\documentclass[aps,prb,twocolumn,floatfix,superscriptaddress,showpacs]{revtex4-1}

\usepackage{graphicx}
\usepackage{amsfonts}
\usepackage{amsmath}
\usepackage{amssymb}
\usepackage{bm}
\usepackage{color}
\usepackage[switch]{lineno}
\usepackage[hypertex]{hyperref}

\begin{document}
\title{Long-range supercurrents induced by the interference effect of opposite-spin triplet state in clean superconductor-ferromagnet structures}
\author{Hao Meng}
\affiliation{Department of Physics, South University of Science and Technology of China, Shenzhen, 518055, China}
\affiliation{School of Physics and Telecommunication Engineering, Shaanxi University of Technology, Hanzhong 723001, China}
\affiliation{Shanghai Key Laboratory of High Temperature Superconductors, Shanghai University, Shanghai 200444, China}
\author{Jiansheng Wu}
\email{wu.js@sustc.edu.cn}
\affiliation{Department of Physics, South University of Science and Technology of China, Shenzhen, 518055, China}
\author{Xiuqiang Wu}
\affiliation{National Laboratory of Solid State Microstructures and Department of Physics, Nanjing University, Nanjing 210093, China}
\author{Mengyuan Ren}
\affiliation{School of material science and technology, Harbin university of science and technology, Harbin 150080, China}
\author{Yajie Ren}
\affiliation{School of Physics and Telecommunication Engineering, Shaanxi University of Technology, Hanzhong 723001, China}
\date{\today}

   \begin{abstract}

   By now it is known that in an s-wave superconductor-ferromagnet-superconductor ($SFS$) structure the supercurrent induced by spin singlet pairs can only transmit a short distance of the order of magnetic coherence length. The long-range supercurrent, taking place on the length scale of the normal metal coherence length, will be maintained by equal-spin triplet pairs, which can be generated by magnetic inhomogeneities in the system. In this paper, we show an unusual long-range supercurrent, which can take place in clean $SF_1F_2S$ junction with non-parallel orientation of magnetic moments. The mechanism behind the enhancement of Josephson current is provided by the interference of the opposite-spin triplet states derived from $S/F_1$ and $F_2/S$ interfaces when both ferromagnetic layers have the same values of the length and exchange field. This finds can provide a natural explanation for recent experiment [Robinson \emph{et al.}, Phys. Rev. Lett. \textbf{104}, 207001 (2010)].

  \end{abstract}

   \pacs{74.45.+c, 74.78.Fk, 73.20.At, 73.40.-c} \maketitle

  \section {Introduction}
  The interplay between superconductivity and ferromagnetism in hybrid structures has currently attracted considerable attention because of the rich unusual physical phenomena~\cite{ZoharNussinov,IverBSperstad,MadalinaColci,KueiSun} and potential practical applications~\cite{AAGolubov,Buz,BerRMP,Esc}. Much effort has been devoted to obtaining a better understanding of the exotic phenomena appeared in heterostructures involving superconductor ($S$) and ferromagnet ($F$). To mention a few of these, it is natural to highlight the experimental and theoretical study of the transport properties in $SF$ heterostructures.

  When a conventional s-wave $S$ is adjacent to a homogeneous $F$, the superconducting proximity effect in this $F$ is rather short ranged due to the differential action of the ferromagnetic exchange field acting on the spin-up and spin-down electrons that form a Cooper pair. In this case, the spin-split of the electronic energy bands in the ferromagnetic region will make the opposite-spin Cooper pair acquire a finite center-of-mass momentum $Q=2h_0/\hbar{v_F}$, where $\emph{h}_0$ and $v_F$ are the exchange field strength and the Fermi velocity, respectively. As a result, the Cooper pair $\mid\uparrow\downarrow\rangle{e^{iQ\cdot{R}}}-$$\mid\downarrow\uparrow\rangle{e^{-iQ\cdot{R}}}$ can be decomposed into a spin singlet component $(\mid\uparrow\downarrow\rangle-$$\mid\downarrow\uparrow\rangle)\cos(Q\cdot{R})$ and a spin triplet component with zero spin projection along the magnetization axis $i(\mid\uparrow\downarrow\rangle+$$\mid\downarrow\uparrow\rangle)\sin(Q\cdot{R})$, where $R$ is the distance from the $S/F$ interface. For simplicity, we will hereafter refer to the wave function of this triplet component as opposite-spin triplet state. Accordingly, the above singlet and triplet components are short range and decays at a distance $\xi_f$ from the superconductor~\cite{Buz,BerRMP}. Here $\xi_f$ is the superconducting coherence length in the $F$ layer, which is much smaller than the correlation length $\xi_n$ in normal metal ($N$). Another peculiarity in such systems is the spatial oscillations of these two components inside the $F$ region~\cite{MatthiasEschrig}. Owing to this oscillatory nature, the critical current of $SFS$ junctions becomes an oscillating function of the $F$ layer thickness. This oscillating behavior of the supercurrent corresponds to the transition between so-called ``0 state¡± and ¡°$\pi$ state¡±~\cite{AAGolubov,Buz}.

  In contrast, it is useful to seek ways to enhance the proximity effect. Several options have recently been proposed in the literature. First, the presence of the inhomogeneous magnetization may strongly modify the $SF$ proximity effect~\cite{Esc,MatthiasEschrig}. In the presence of domain at the $S/F$ interface, the induced spin triplet pairing with the equal spin projection $\mid\uparrow\uparrow\rangle$ or $\mid\downarrow\downarrow\rangle$ can propagate long distances in a ferromagnetic material. The primary reason is that since two triplet-paired electrons at the Fermi surface have no momentum difference and propagate with the same phase, they are not affected by the exchange field and decay at a distance $\xi_n$. This long-range proximity effect, giving rise to induced superconducting correlations in ferromagnets and half-metals, is prime examples of the potential that lies within this field of research. It has been observed in Co~\cite{TruptiSKhaire,CarolinKlose,JWARobinson} and in the half-metal CrO$_2$~\cite{RSKeizer,MSAnwar}. Its origin is related with the presence of the spin-flip scattering at the $S/F$ interface, which is induced by the non-collinear magnetic domain or magnetic impurity.

  Recently, the second way to enhance the supercurrent has been proposed in $SFS$ junction containing a noncollinear thin magnetic domain in the center of ferromagnetic region~\cite{HM,AVSamokhvalov}. The magnetic domain will induce a spin-flip scattering process, which reverses the spin orientations of the singlet Cooper pair and simultaneously changes the sign of the corresponding electronic momentum. Under these conditions the singlet Cooper pair will create an exact phase-cancellation effect and gets an additional $\pi$ phase shift as it passes through the entire ferromagnetic region, so that the supercurrent can not be suppressed.

  The third approach requires the magnetizations in the clean $SF_1F_2S$ junction to be arranged antiparallel. This situation was previously proposed by Blanter \emph{et al.}~\cite{YaMBlanter} through solving the Eilenberger equation. However, the physical origin of this enhanced proximity effect is more subtle. With the simplest picture of this situation, the authors argue that when the Cooper pair propagating from the first $F$ layer to the second between the superconducting electrodes, it first acquire a relative phase $\delta\varphi_1=Q\cdot{R_1}$, where $R_1$ is the distance traversed in the first ferromagnetic layer. Subsequently, in the second layer with opposite direction of exchange field, the above pair will gain the other phase $\delta\varphi_2=-Q\cdot{R_2}$, which can partially compensate for $\delta\varphi_1$. For $R_1=R_2$ they have full compensation, then the ferromagnetic bilayer behaves as a piece of normal metal, and the proximity effect is fully restored. However, this explanation dose not specify which pairing form ($\mid\uparrow\downarrow\rangle-$$\mid\downarrow\uparrow\rangle$ or $\mid\uparrow\downarrow\rangle+$$\mid\downarrow\uparrow\rangle$) provides the main contribution to the long-range Josephson current. Soon afterwards, the same conclusion for clean junction was proposed theoretically by Pajovi$\acute{c}$ \emph{et al.}~\cite{ZPajovic} via solving the Bogoliubov-de Gennes (BdG) equation, but they just took into account a single transverse channel case for simplicity, which is inconsistent with the realistic situation.  Recently, Robinson \emph{et al.}~\cite{Robinson} observed experimentally that the supercurrent in the antiparallel domain configuration was enhanced with respect to the parallel one.

  \begin{figure}
  \centerline{\includegraphics[width=3.8in]{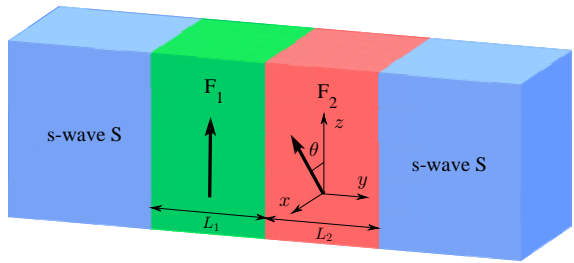}} 
  \caption{(color online) The Josephson junction consists of two s-wave superconductors and two ferromagnets of the thicknesses $L_1$ and $L_2$. The exchange fields of the ferromagnets, $\vec{h}_1$ and $\vec{h}_2$, denoted by the thick arrows, are confined to the $x$-$z$ plane, but are misaligned by an angle $\theta$. The phase difference between the two superconductors is $\phi$$=$$\phi_R$$-$$\phi_L$.}
  \label{Fig.1}
  \end{figure}

  In this paper, we report a manifestation of the interference effect in clean $SF_1F_2S$ junction with non-parallel magnetizations by considering an oblique injection process. Note that in contrast to the model of Ref.~\cite{ZPajovic} we consider the multiple transverse channel that is more agree with the realistic case of the planar junctions. We investigate the dependence of the critical Josephson current on the thicknesses of both $F$ layers. It is shown a slowly decaying characteristic for non-parallel orientation of magnetizations in the $F$ layers. Furthermore, by changing the relative magnetization direction of $F$ layers from parallel to antiparallel, the critical current is varied from a small to a large value. In this process, the spin singlet state changes slightly but the opposite-spin triplet state could switch from a finite value to be cancelled out in central region of the entire $F$ layer. So we attribute the enhancement of critical current to the interference effect of opposite-spin triplet wave functions in the $F$ region. This effect can weaken the role of the center-of-mass momentum acquired by the Cooper pair, and the situation is similar to the transition of the Cooper pair in normal metal, in which case only the singlet state exists but the opposite-spin triplet state disappear. Moreover, it is found that the critical current is inversely proportional to the exchange field of both $F$ layers. When two $F$ layers are converted into half-metals the Joseohson current will be prohibited. That is because the singlet and triplet states will all be suppressed by the exchange splitting of two $F$ layers, and the interference effect could completely vanish in the entire $F$ region. On the other hand, if the both $F$ layers have different features, the critical current will oscillator decay with their difference of the lengths or exchange fields, which can be attribute to the variation of the interference between the two opposite-spin triplet states derived from the $S/F_1$ and $F_2/S$ interfaces.

  \section{Model and formula}
  The $SF_1F_2S$ junction we consider is shown schematically in Fig.~\ref{Fig.1}. We denote the ferromagnetic layer thicknesses by $L_1$ and $L_2$, respectively. The $y$ axis is chosen to be perpendicular to the layer interfaces with the origin at the $S/F_1$ interface, and the whole system satisfies translational invariance in the $x$-$z$ plane. The exchange field in the $F_1$ layer is directed along the $z$ axis while within the $F_2$ layer, it is oriented at an angle $\theta$ in the $x$-$z$ plane.

  The BCS mean-field effective Hamiltonian~\cite{Buz,PGdeGennes} is
  \begin{equation}
  \begin{aligned}
   H_{eff}&=\int{d\vec{r}}\{\sum_{\alpha}\psi^{\dag}_{\alpha}(\vec{r})H_e\psi_{\alpha}(\vec{r})+\frac{1}{2}[\sum_{\alpha,\beta}(i\sigma_{y})_{\alpha\beta}\Delta(\vec{r})\\
   &\psi^{\dag}_{\alpha}(\vec{r})\psi^{\dag}_{\beta}(\vec{r})+h.c.]-\sum_{\alpha,\beta}\psi^{\dag}_{\alpha}(\vec{r})(\vec{h}\cdot\vec{\sigma})_{\alpha\beta}\psi_{\beta}(\vec{r})\}
  \end{aligned}
  \label{Heff}
  \end{equation}
  where $H_e=-\hbar^{2}\nabla^{2}/2m-E_F$, $\psi^{\dag}_{\alpha}(\vec{r})$ and $\psi_{\alpha}(\vec{r})$ represent creation and annihilation operators with spin $\alpha$, and the vector $\vec{\sigma}=(\sigma_x, \sigma_y, \sigma_z)$ is composed of Pauli spin matrices. $m$ is the effective mass of the quasiparticles in both $S$s and $F$s, and $E_F$ is the Fermi energy. $\Delta(\vec{r})=\Delta(T)[e^{i\phi_{L}}\Theta(-y)+e^{i\phi_{R}}\Theta(y-L_{F})]$ describes the superconducting pair potential with $L_F=L_{1}+L_{2}$. Here $\Delta(T)$ accounts for the temperature-dependent energy gap. It satisfies the BCS relation $\Delta(T)=\Delta_0\tanh(1.74\sqrt{T_c/T-1})$, where $\Delta_0$ is the energy gap at zero temperature and $T_c$ is the superconducting critical temperature. $\phi_{L(R)}$ is the phase of the left (right) $S$, and $\Theta(y)$ is the unit step function. The exchange field $\vec{h}$ due to the ferromagnetic magnetizations in the $F$ region can be written as
  $$\vec{h}=
  \begin{cases}
   h_1\hat{z}, & \text{$0<y<L_1$}\\
   h_2(\sin\theta\hat{x}+\cos\theta\hat{z}), & \text{$L_1<y<L_F$}.
  \end{cases}$$
  To diagonalize the effective Hamiltonian, we make use of the Bogoliubov transformation $\psi_{\alpha}(\vec{r})=\sum_{n}[u_{n\alpha}(\vec{r})\hat{\gamma}_{n}+v^{\ast}_{n\alpha}(\vec{r})\hat{\gamma}^{\dag}_{n}]$ and take into account the anticommutation relations of the quasiparticle annihilation operator $\hat{\gamma}_{n}$ and creation operator $\hat{\gamma}^{\dag}_{n}$. The resulting BdG equation can be expressed as~\cite{PGdeGennes}

  \begin{equation}
  \begin{pmatrix}
	\hat{H}(\vec{r}) & i\hat{\sigma}_{y}\Delta(\vec{r}) \\
    -i\hat{\sigma}_{y}\Delta^{\ast}(\vec{r}) & -\hat{H}(\vec{r}) \\
  \end{pmatrix}
  \begin{pmatrix}
	\hat{u}(\vec{r}) \\
    \hat{v}(\vec{r}) \\
  \end{pmatrix}
  =E
  \begin{pmatrix}
	\hat{u}(\vec{r}) \\
    \hat{v}(\vec{r}) \\
  \end{pmatrix},
  \label{BdG}
  \end{equation}
  where $\hat{H}(\vec{r})=H_e\hat{\textbf{1}}-h_z(\vec{r})\hat{\sigma}_{z}-h_{x}(\vec{r})\hat{\sigma}_{x}$ and $\hat{\textbf{1}}$ is the unity matrix. Besides, $\hat{u}(\vec{r})=[u_{\uparrow}(\vec{r}), u_{\downarrow}(\vec{r})]^T$ and $\hat{v}(\vec{r})=[v_{\uparrow}(\vec{r}), v_{\downarrow}(\vec{r})]^{T}$ are quasiparticle and quasihole wave functions, respectively.

  In order to calculate the Josephson current, we adopt the Blonder-Tinkham-Klapwijk (BTK) approach. The BdG equation (\ref{BdG}) can be solved for each superconducting electrode and each ferromagnetic layer, respectively. We have four different incoming quasiparticles, electronlike quasiparticles (ELQs) and holelike quasiparticles (HLQs) with spin-up and spin-down. For an incident spin-up electron in the left superconducting electrode, the wave function is
  \begin{equation}
  \begin{aligned}
  \Psi^{S}_{L}(y)&=\hat{M}_1e^{ik_{e}y}+a_1\hat{M}_2e^{ik_{h}y}+b_1\hat{M}_1e^{-ik_{e}y} \\
                 &+a'_1\hat{M}_3e^{ik_{h}y}+b'_1\hat{M}_4e^{-ik_{e}y}.
  \end{aligned}
  \label{SL}
  \end{equation}
  In this particular process, the coefficients $b_{1}$, $b'_{1}$, $a'_{1}$, and $a_{1}$ correspond to the normal reflection, the normal reflection with spin-flip, the novel Andreev reflection, and the usual Andreev reflection, respectively. Moreover, $\hat{M}_1=[ue^{i\phi_{L}/2}, 0, 0, ve^{-i\phi_{L}/2}]^{T}$, $\hat{M}_2=[ve^{i\phi_{L}/2}, 0, 0, ue^{-i\phi_{L}/2}]^{T}$, $\hat{M}_3=[0, -ve^{i\phi_{L}/2}, ue^{-i\phi_{L}/2}, 0]^{T}$, and $\hat{M}_4=[0, ue^{i\phi_{L}/2}, -ve^{-i\phi_{L}/2}, 0]^{T}$ are the four basis wave functions of the left $S$, in which the quasiparticle amplitudes are defined as $u=\sqrt{(1+\Omega/E)/2}$, $v=\sqrt{(1-\Omega/E)/2}$ and $\Omega=\sqrt{E^2-\Delta^2}$. $k_{e(h)}=\sqrt{2m[E_F+(-)\Omega]/\hbar^2-k^{2}_{\parallel}}$ are the perpendicular components of the ELQs (HLQs) wave vector with $k_{\parallel}$ as the parallel component.

  The corresponding wave function in the right superconducting electrode is
  \begin{equation}
  \begin{aligned}
  \Psi^{S}_{R}(y)&=c_1\hat{N}_1e^{ik_{e}y}+d_1\hat{N}_2e^{-ik_{h}y}+c'_1\hat{N}_4e^{ik_{e}y} \\
                 &+d'_1\hat{N}_3e^{-ik_{h}y},
  \end{aligned}
  \label{SR}
  \end{equation}
  where the transmission coefficients $c_1$, $d_1$, $c'_1$, and $d'_1$ correspond to the reflection processes described above. The basis wavefunctions $\hat{N}_p$ ($p=1\text{--}4$) in the right $S$ can be obtained from $\hat{M}_p$ by performing the substitution $\phi_L\rightarrow\phi_R$.

  The wave function in the $F_2$ layer can be described by transformation matrix~\cite{LiJingJin} as
  \begin{equation}
  \begin{aligned}
  \Psi^{F}_{2}(y)&=\hat{T}\{[e\cdot{exp(ik^{e\uparrow}_{F2}y)}+f\cdot{exp(-ik^{e\uparrow}_{F2}y)}]\hat{e}_{1} \\
  &+[e'\cdot{exp(ik^{e\downarrow}_{F2}y)}+f'\cdot{exp(-ik^{e\downarrow}_{F2}y)}]\hat{e}_{2} \\
  &+[g\cdot{exp(-ik^{h\uparrow}_{F2}y)}+h\cdot{exp(ik^{h\uparrow}_{F2}y)}]\hat{e}_{3} \\
  &+[g'\cdot{exp(-ik^{h\downarrow}_{F2}y)}+h'\cdot{exp(ik^{h\downarrow}_{F2}y)}]\hat{e}_{4}\}.
  \end{aligned}
  \label{F2}
  \end{equation}
  Here $\hat{e}_{1}=[1,0,0,0]^{T}$, $\hat{e}_{2}=[0,1,0,0]^{T}$, $\hat{e}_{3}=[0,0,1,0]^{T}$, $\hat{e}_{4}=[0,0,0,1]^{T}$ are basis wave functions in the ferromagnetic region, and $k^{e(h)\alpha}_{F2}=\sqrt{2m[E_F+(-)E+\rho_{\alpha}h_{2}]/\hbar^2-k^{2}_{\parallel}}$ with $\rho_{\uparrow(\downarrow)}=1(-1)$ are the perpendicular components of wave vectors for ELQs and HLQs. It is worthy to note that the parallel component $k_{\parallel}$ is conserved in transport processes of the quasiparticles. The transformation matrix has been defined as $\hat{T}=\hat{\textbf{1}}\otimes(\cos\frac{\theta}{2}\cdot\hat{\textbf{1}}-i\cdot{\sin\frac{\theta}{2}}\cdot\hat{\sigma}_{y})$. From the conversion $\theta\rightarrow0$ and $h_2\rightarrow{h_1}$, we can obtain the wave function $\Psi^{F}_1(y)$ in the $F_1$ layer.

  All scattering coefficients can be obtained by continuity of the wave functions and their derivatives at the interfaces:
  \begin{equation}
  \begin{aligned}
  &\Psi^{S}_L(y_1)=\Psi^{F}_1(y_1),\partial_{y}[\psi^{F}_{1}-\psi^{S}_{L}]|_{y_1}=2k_FZ_1\psi^{F}_{1}(y_1);\\
  &\Psi^{F}_{1}(y_2)=\Psi^{F}_2(y_{2}),\partial_{y}[\psi^{F}_{2}-\psi^{F}_{1}]|_{y_{2}}=2k_FZ_2\psi^{F}_{2}(y_{2});\\
  &\Psi^{F}_{2}(y_3)=\Psi^{S}_R(y_3),\partial_{y}[\psi^{S}_{R}-\psi^{F}_{2}]|_{y_3}=2k_FZ_3\psi^{S}_{R}(y_3).
  \end{aligned}
  \label{WF}
  \end{equation}
  Here, $Z_1\text{--}Z_3$ are dimensionless parameters describing the magnitude of the interfacial resistances. $y_{1\text{--}3}=0, L_1, L_F$ are local coordinate values at the layer interfaces, and $k_F=\sqrt{2mE_F}$ is the Fermi wave vector. The wave functions for the other types of quasiparticle injection processes can be obtained in a similar way. From the boundary conditions, we obtain a system of linear equations that yield the scattering coefficients. With these coefficients at hand, we can use the finite-temperature Green's function formalism~\cite{AFurusaki,ZMZheng,YTanaka} to calculate dc Josephson current,
  \begin{equation}
  \begin{aligned}
  &I_{e}(\phi)=\frac{k_BTe\Delta}{4\hbar}\sum_{k_{\parallel}}\sum_{\omega_{n}}\frac{k_{e}(\omega_{n})+k_{h}(\omega_{n})}{\Omega_{n}} \\
  &\times[\frac{a_1(\omega_{n},\phi)-a_2(\omega_{n},\phi)}{k_{e}}+\frac{a_3(\omega_{n},\phi)-a_4(\omega_{n},\phi)}{k_{h}}],
  \end{aligned}
  \label{Ie}
  \end{equation}
  where $\omega_{n}=\pi{k_B}T(2n+1)$ are the Matsubara frequencies with $n=0, 1, 2,...$ and $\Omega_{n}=\sqrt{\omega^{2}_{n}+\Delta^{2}(T)}$. $k_e(\omega_{n})$, $k_h(\omega_{n})$, and $a_j(\omega_{n},\phi)$ with $j=1, 2, 3, 4$ are obtained  from $k_e$, $k_h$, and $a_j$ by analytic continuation $E\rightarrow{i}\omega_{n}$. In this case the critical current is defined by $I_c=max_{\phi}|I_e(\phi)|$.

  To acquire the time dependent triplet amplitude functions and the local density of the states (LDOS), we solve the BdG equation~(\ref{BdG}) by Bogoliubov's self-consistent field method~\cite{PGdeGennes,JBKetterson,KlausHalterman,HaoMeng}. The $SF_1F_2S$ junction is placed in a one-dimensional square potential well with infinitely high walls, then the eigenvalues and eigenvectors of the equation~(\ref{BdG}) have the following substitutions: $E\rightarrow{E_n}$ and $[u_{\uparrow}(\vec{r}), u_{\downarrow}(\vec{r}), v_{\uparrow}(\vec{r}), v_{\downarrow}(\vec{r})]^{T}\rightarrow[u_{n\uparrow}(\vec{r}), u_{n\downarrow}(\vec{r}), v_{n\uparrow}(\vec{r}), v_{n\downarrow}(\vec{r})]^{T}$. Accordingly, the corresponding quasiparticle amplitudes can be expanded in terms of a set of basis vectors of the stationary states~\cite{LDLandau}, $u_{n\alpha}(\vec{r})=\sum_{q}u^{\alpha}_{nq}\zeta_{q}(y)$ and $v_{n\alpha}(\vec{r})=\sum_{q}v^{\alpha}_{nq}\zeta_{q}(y)$ with $\zeta_{q}(y)=\sqrt{2/L}\sin(q{\pi}y/L)$. Here $q$ is a positive integer and $L=L_{S1}+L_F+L_{S2}$, where $L_{S1}$ and $L_{S2}$ are thickness of left and right superconductors, respectively. The pair potential in the BdG equation~(\ref{BdG}) satisfies the self-consistency condition~\cite{PGdeGennes}
  \begin{equation}
  \begin{aligned}
  \Delta(y)&=\frac{g(y)}{2}\sum_{n}{'}\sum_{qq'}(u_{nq}^{\uparrow}v_{nq'}^{\downarrow*}-u_{nq}^{\downarrow}v^{\uparrow*}_{nq'})\zeta_{q}(y)\zeta_{q'}(y)\\
           &\times\tanh(\frac{E_n}{2k_BT}),
  \end{aligned}
  \label{Det}
  \end{equation}
  where the primed sum of $E_n$ is over eigenstates corresponding to positive energies smaller than or equal to the Debye cutoff energy $\omega_{D}$, and the superconducting coupling parameter $g(y)$ is a constant in the superconducting regions and zero elsewhere. The BdG equation~(\ref{BdG}) is solved by an iterative schedule. One first starts from the stepwise approximation for the pair potential and iterations are performed until the change in value obtained for $\Delta(y)$ does not exceed a small threshold value. The amplitude functions of the spin triplet state with zero and net spin projection are defined, respectively, as follows~\cite{KlausHalterman}
  \begin{equation}
   f_{0}(y,t)=\frac{1}{2}\sum_{n}\sum_{qq'}(u^{\uparrow}_{nq}v^{\downarrow*}_{nq'}+u^{\downarrow}_{nq}v^{\uparrow*}_{nq'})\zeta_{q}(y)\zeta_{q'}(y)\eta_n(t),
  \label{F0}
  \end{equation}
  \begin{equation}
   f_{1}(y,t)=\frac{1}{2}\sum_{n}\sum_{qq'}(u_{nq}^{\uparrow}v^{\uparrow*}_{nq'}-u_{nq}^{\downarrow}v^{\downarrow*}_{nq})\zeta_{q}(y)\zeta_{q'}(y)\eta_n(t),
  \label{F1}
  \end{equation}
  where the sum of $E_n$ is in general performed over all positive energies, and $\eta_n(t)=\cos(E_nt)-i\sin(E_nt)\tanh(E_n/2k_BT)$. Additionally, the amplitude function of the spin singlet state can be written as $f_3\equiv\Delta(y)/g(y)$. In this paper the singlet and triplet amplitude functions are all normalized to the value of the singlet pairing amplitude in a bulk superconducting material. The LDOS is given by~\cite{KlausHalterman}
  \begin {equation}
  \begin{aligned}
   N(y,\epsilon)=&-\sum_{n}{'}\sum_{qq'}[(u_{nq}^{\uparrow}u^{\uparrow*}_{nq'}+u_{nq}^{\downarrow}u^{\downarrow*}_{nq'})f'(\epsilon-E_n)\\
                 &+(v_{nq}^{\uparrow}v^{\uparrow*}_{nq'}+v_{nq}^{\downarrow}v^{\downarrow*}_{nq'})f'(\epsilon+E_n)]\zeta_{q}(y)\zeta_{q'}(y),
  \end{aligned}
  \label{LDOS}
  \end {equation}
  where $f'(\varepsilon)=\partial{f}/\partial{\varepsilon}$ is the derivative of the Fermi function. The LDOS is normalized by its value at $\epsilon=3\Delta_0$ beyond which LDOS is almost constant.

  \section{Results and Discussions}

  Unless otherwise stated, in BTK approach we use the superconducting gap $\Delta_0$ as the unit of energy. The Fermi energy is defined as $E_F=1000\Delta_0$, and the temperature is taken to be $T/T_c=0.1$. We assume all interfaces between the layers are transparent for electrons $Z_{1-3}=0$. All lengths and the exchange field strengths are measured in units of the inverse of the Fermi wave vector $k_F$ and the Fermi energy $E_F$, respectively. In Bogoliubov's self-consistent field method, we consider the low-temperature limit and set $k_FL_{S1}=k_FL_{S2}=400$ and $\omega_{D}/E_F=0.1$, the other parameters are the same as the ones described above.

  \begin{figure}
  \includegraphics[width=3.3in]{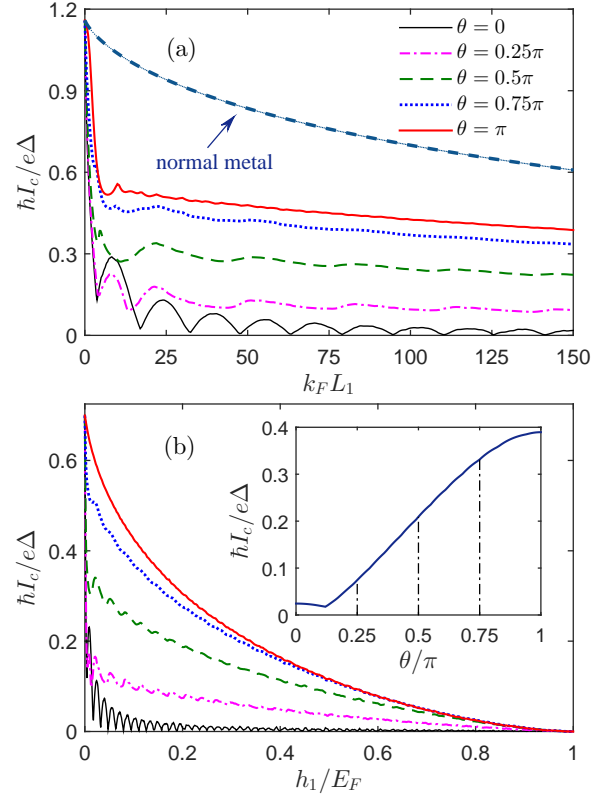} 
  \caption{(Color online) (a) Critical current as a function of thickness $k_FL_1$(=$k_FL_2$) for exchange field $h_1/E_F$=$h_2/E_F$ =0.1. Several misorientation angles $\theta$ are considered as depicted in the legend. (b) Critical current as a function of $h_1/E_F$(=$h_2/E_F$) for $k_FL_1$=$k_FL_2$=100. The inset depicts $I_c$ versus $\theta$ for $k_FL_1$=$k_FL_2$=100 and $h_1/E_F$=$h_2/E_F$=0.1. Two main panels utilize the same legend.}
  \label{Fig.2}
  \end{figure}

  The detailed dependence of the critical current on the thickness $k_FL_1(=k_FL_2)$ is shown in Fig.~\ref{Fig.2}(a) for different misorientation angles $\theta$. We can find a significant change in the magnitude of critical current depending on the mutual orientation of two ferromagnetic magnetizations. Considering first the parallel orientation ($\theta=0$), the well know $0$-$\pi$ oscillations are reproduced, where the current change sign for certain values of thickness. It should however be noted that we have taken absolute value for $I_e(\phi)$ to define the critical current $I_c$, because that is most commonly measured in experiments. Increasing the misorientation angle $\theta$ tends to enhance the amplitude of current. Meanwhile, the oscillations of the critical current with ferromagnetic layer thickness will diminish. For perpendicular case ($\theta=0.5\pi$), the oscillations will almost cease, leaving the junction in the 0 state for larger values of $k_FL_1$. In addition, we can observe a clear maximum of the critical current for an antiparallel magnetizations ($\theta=\pi$), but it is significantly smaller than that in $SNS$ junction for all values of $k_FL_1$. This conclusion is inconsistent with the previous results of Ref.~\cite{ZPajovic,YaMBlanter}.

  \begin{figure}
  \centerline{\includegraphics[width=3.1in]{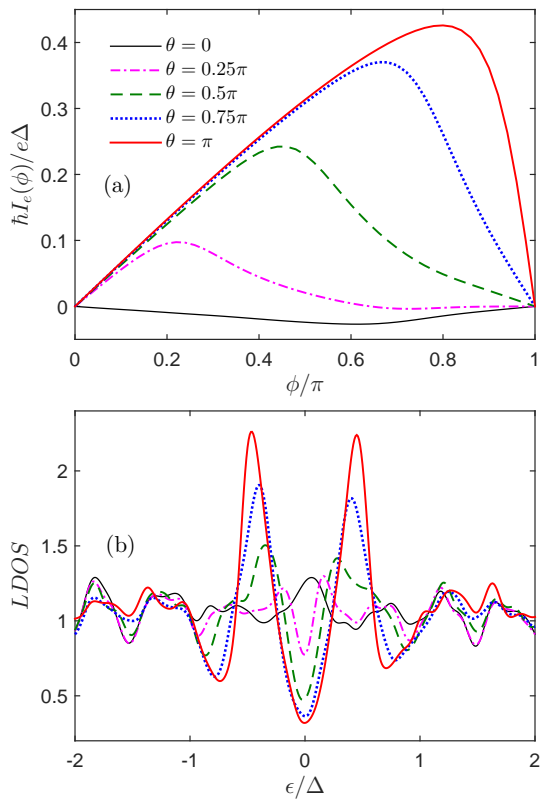}} 
  \caption{(Color online) (a) The current-phase relation $I_e(\phi)$ and (b) the LDOS in the center of $F$ layer ($k_Fy$=100) for several values of the misorientation angle $\theta$. The LDOS is calculated at $k_BT$=0.0008. Parameters used in all panels are $k_FL_1$=$k_FL_2$=100 and $h_1/E_F$=$h_2/E_F$=0.1.}
  \label{Fig.3}
  \end{figure}

  By comparison, the dependence of critical current $I_c$ on the exchange field $h_1/E_F(=h_2/E_F)$ is plotted in Fig.~\ref{Fig.2}(b). It can be clearly seen that for various $\theta$ the critical current $I_c$ decreases monotonically with increasing $h_1/E_F$ and it decreases down to zero at $h_1/E_F=1$, which suggests a vanishing of the Josephson current. This phenomenon shows that the strong exchange splitting of the energy bands inside the $F$ layers could effectively damp the tunneling of pairing electrons. For $\theta=0$, the critical current becomes an oscillating function of the $h_1/E_F$, and is also accompanied by an exponential decay. This oscillating effect will diminish as the enhancement of $\theta$ and also disappear at some larger $\theta$. We confirm the obvious fact that the critical current increases with $\theta$ for any fixed $h_1/E_F$. Inset of Fig.~\ref{Fig.2}(b) shows this character of critical current for $h_1/E_F=0.1$. It displays a nonmonotonic dependence of the critical current on $\theta$, where a low dip corresponds to $\theta=0.12\pi$ and the maximum is located at $\theta=\pi$. The main reason is because the junction starts out in the $\pi$ state for the parallel orientation, and we can see that a transition from the $\pi$ state to the 0 state takes place as the increase of $\theta$. In contrast, if the 0 state is the equilibrium state of the junction for $\theta=0$, we will acquire a monotonic variation of the critical current when $\theta$ varies from 0 to $\pi$. These behaviors agree with the statement made in Ref.~\cite{IverBSperstad,YuSBarash}.

  \begin{figure*}
  \includegraphics[width=6.2in]{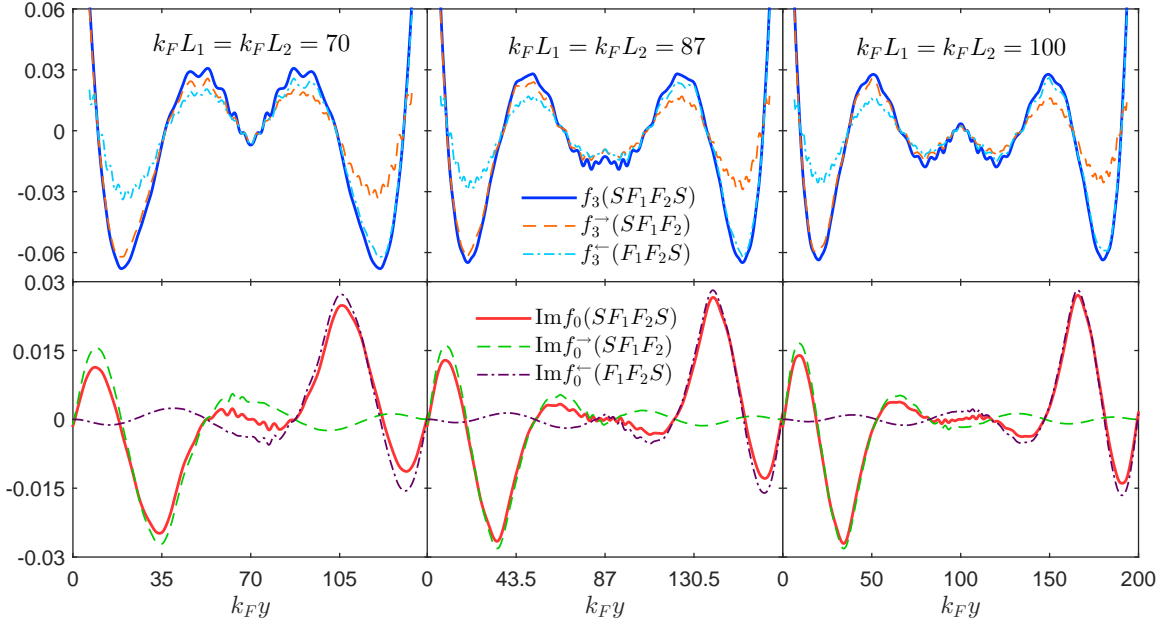} 
  \caption{(Color online) The singlet components (top row) and the imaginary parts of opposite-spin triplet components (bottom row) plotted as a function of the coordinate $k_Fy$ for three values of lengths $k_FL_1$=$k_FL_2$=70 (left column), 87 (middle column) and 100 (right column) in the antiparallel magnetizations. Here $f_{3(0)}$, $f_{3(0)}^{\rightarrow}$ and $f_{3(0)}^{\leftarrow}$ correspond to $SF_1F_2S$, $SF_1F_2$ and $F_1F_2S$ configurations, respectively. Parameters used in all panels are $h_1/E_F$=$h_2/E_F$=0.1, $\theta$=$\pi$, $\omega_Dt$=4, and $\phi$=0.}
  \label{Fig.4}
  \end{figure*}

  In order to clearly illustrate above feature of the critical current, we plot the current-phase relation $I_e(\phi)$ and the LDOS respectively in Figs.~\ref{Fig.3}(a) and \ref{Fig.3}(b) for several misorientation angles $\theta$. If the two ferromagnetic layers have the same directions ($\theta=0$), the Josephson current $I_e(\phi)$ is negative and its amplitude is small enough, then the LDOS displays a very small conductance peak at the Fermi level ($\epsilon=0$), as plotted in Fig.~\ref{Fig.3}(b). These features indicate the junction is situated in $\pi$ state. By contrast, the current will turn to positive quantity and its amplitude is correspondingly enhanced by increasing the misorientation angle $\theta$. Under such circumstances, the LDOS at $\epsilon=0$ will be turned from peak to valley. When $\theta$ increase to $\pi$, the LDOS is strongly enhanced with two distinguishable peaks nearly at $\epsilon=\pm0.5\Delta$. Such LDOS shapes represent the ground state of the junction is converted into the 0 state. These behaviors demonstrate that the transition between the $\pi$ state and 0 state can be realized by tuning the relative orientation of magnetizations for an appropriate ferromagnetic thicknesses.

  For searching the main reason of enhancement of the critical current, we first focus on the transmission of the singlet and triplet components in the antiparallel orientation of magnetic moments. In Fig.~\ref{Fig.4}, we show the spatial distribution of the singlet component and the imaginary parts of the opposite-spin triplet component for three different lengths $k_FL_1=k_FL_2=70$, 87 and 100. In the panels, $f_{3(0)}$, $f_{3(0)}^{\rightarrow}$ and $f_{3(0)}^{\leftarrow}$ represent the wave functions in $SF_1F_2S$, $SF_1F_2$ and $F_1F_2S$ configurations, respectively. It is found that the singlet components $f_3$ are symmetrical about the $F_1/F_2$ interface, but the triplet components $f_0$ are antisymmetric and their amplitudes will diminish nearly at the central region of the $F$ layer. The physical origin of these effects can be described as follows. Due to the exchange splitting, the original Cooper pair $\mid\uparrow\downarrow\rangle-$$\mid\downarrow\uparrow\rangle$ in the left superconducting electrode will acquire a center-of-mass momentum $Q$ in the $F_1$ region, then this pair can be transformed into $\mid\uparrow\downarrow\rangle{e^{iQ\cdot{R}}}-$$\mid\downarrow\uparrow\rangle{e^{-iQ\cdot{R}}}$, where $R$ represents the transmission distance from the $S/F_1$ interface. Accordingly, the wave function of the Cooper pair can be rewritten as a mixture of the singlet component and the opposite-spin triplet component: ($\mid\uparrow\downarrow\rangle-$$\mid\downarrow\uparrow\rangle$)$\cos$($Q\cdot{R}$)+$i\cdot$($\mid\uparrow\downarrow\rangle$+$\mid\downarrow\uparrow\rangle$)$\sin$($Q\cdot{R}$). Additionally, for the antiparallel magnetic moment the wave-vector mismatches for spin-up and spin-down particles at both side of the $F_1/F_2$ interface will result in an interface scattering~\cite{MEschrigTL}. The right-going particle wave transmitted from the $F_1$ layer will take the $F_1/F_2$ interface as the wave source and continually transports into the $F_2$ layer. In addition, at the location of $F_1/F_2$ interface the phase of the wave function could maintain continuously in above transmission process, but the center-of-mass momentum $Q$ will be transformed into $-Q$ in the $F_2$ layer. As a result, the right-gong wave function of the Cooper pair arising from the $S/F_1$ interface can be written as
  $$\chi^{\rightarrow}=
  \begin{cases}
  \mid\uparrow\downarrow\rangle{e^{iQ\cdot{R_r}}}-\mid\downarrow\uparrow\rangle{e^{-iQ\cdot{R_r}}},  & \text{in $F_1$ layer} \\
  \mid\uparrow\downarrow\rangle{e^{iQ\cdot(L_1-R'_r)}}-\mid\downarrow\uparrow\rangle{e^{-iQ\cdot(L_1-R'_r)}}, & \text{in $F_2$ layer},
  \end{cases}$$
  where $R_r$ and $R'_r$ denote the distance from the $S/F_1$ and $F_1/F_2$ interfaces, respectively. This wave function can be decomposed into the singlet and triplet components. Accordingly, the right-going singlet component is given by
  $$f_{3}^{\rightarrow}=
  \begin{cases}
  (\mid\uparrow\downarrow\rangle-\mid\downarrow\uparrow\rangle)\cos(QR_r),  & \text{in $F_1$ layer} \\
  (\mid\uparrow\downarrow\rangle-\mid\downarrow\uparrow\rangle)\cos[Q(L_1-R'_r)],  & \text{in $F_2$ layer}.
  \end{cases}$$
  And the associated right-going triplet component reads as
  $$f_{0}^{\rightarrow}=
  \begin{cases}
  i(\mid\uparrow\downarrow\rangle+\mid\downarrow\uparrow\rangle)\sin(Q{R_r}),  & \text{in $F_1$ layer} \\
  i(\mid\uparrow\downarrow\rangle+\mid\downarrow\uparrow\rangle)\sin[Q(L_1-R'_r)],  & \text{in $F_2$ layer}.
  \end{cases}$$
  From above descriptions, we can demonstrate that the $f_3^{\rightarrow}$ and $f_0^{\rightarrow}$ are all are symmetrical about the $F_1/F_2$ interface.

  On the other hand, the left-going wave function $\chi^{\leftarrow}$ has the same transmission characteristic, but the only difference is that it generates at the $F_2/S$ interface, in which case its original center-of-mass momentum will become $-Q$ in the $F_2$ region. It can be expressed as
  $$\chi^{\leftarrow}=
  \begin{cases}
  \mid\uparrow\downarrow\rangle{e^{iQ\cdot(R_l-L_2)}}-\mid\downarrow\uparrow\rangle{e^{-iQ\cdot(R_l-L_2)}},  & \text{in $F_1$ layer} \\
  \mid\uparrow\downarrow\rangle{e^{-iQ\cdot{R'_l}}}-\mid\downarrow\uparrow\rangle{e^{iQ\cdot{R'_l}}}, & \text{in $F_2$ layer},
  \end{cases}$$
  where $R_l$ and $R'_l$ represent the distance from the $F_1/F_2$ and $F_2/S$ interfaces, respectively. Hence we can get the left-going singlet component
  $$f_{3}^{\leftarrow}=
  \begin{cases}
  (\mid\uparrow\downarrow\rangle-\mid\downarrow\uparrow\rangle)\cos[Q(R_l-L_2)],  & \text{in $F_1$ layer} \\
  (\mid\uparrow\downarrow\rangle-\mid\downarrow\uparrow\rangle)\cos(QR'_l),  & \text{in $F_2$ layer}
  \end{cases}$$
  and the left-going triplet component
  $$f_{0}^{\leftarrow}=
  \begin{cases}
  i(\mid\uparrow\downarrow\rangle+\mid\downarrow\uparrow\rangle)\sin[Q(R_l-L_2)],  & \text{in $F_1$ layer} \\
  -i(\mid\uparrow\downarrow\rangle+\mid\downarrow\uparrow\rangle)\sin(QR'_l),  & \text{in $F_2$ layer}.
  \end{cases}$$

   From above equations, we can find that because the factor $\cos(QR'_l)$ of the singlet component $f_3^{\leftarrow}$ is an even function of center-of-mass momentum, $f_3^{\leftarrow}$ will not change its sign when passing from the $F_2$ layer into the $F_1$ layer, then it will overlap with $f_3^{\rightarrow}$. In contrast, the triplet component $f_0^{\leftarrow}$ will be added a negative sign because the factor $\sin(QR'_l)$ of this component is an odd function of center-of-mass momentum. Consequently, the sign of $f_0^{\leftarrow}$ is opposite to $f_0^{\rightarrow}$, and these two components could be cancelled out each other. In addition, it is known that in normal metal the singlet component decays more slowly and the triplet component does not exist, then the supercurrent could transmit a long distance in the $SNS$ junction. Compared with this situation, the long-range Josephson current could be induced in the $SF_1F_2S$ junction with antiparallel magnetizations by the interference effect, which can revise the configurations of the singlet and triplet components and make their characters more close to them in the normal metal. In Fig.~\ref{Fig.4}, we show the numerical results about the singlet and triplet components through solving the BdG equation~(\ref{BdG}), which further demonstrate our above discussions. In this case, the total $f_3$ will be enhanced by the coherent superposition of $f_3^{\rightarrow}$ and $f_3^{\leftarrow}$, but $f_0$ will be cancelled out in the cental region of the $F$ layer due to the opposite signs of $f_0^{\rightarrow}$ and $f_0^{\leftarrow}$.

  \begin{figure}
  \includegraphics[width=3.4in]{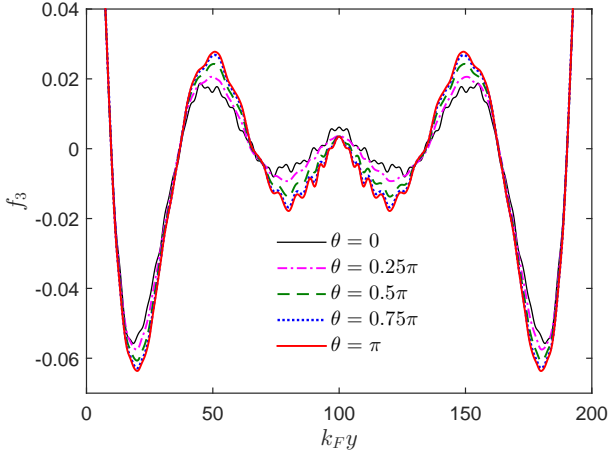} 
  \caption{(Color online) The singlet component $f_3$ plotted as a function of coordinate $k_Fy$ for several values of the misorientation angle $\theta$ as indicated in the legend. The results plotted are for $k_FL_1$=$k_FL_2$=100, $h_1/E_F$=$h_2/E_F$=0.1, and $\phi$=0.}
  \label{Fig.5}
  \end{figure}

  In the following, we want to known which components can make a crucial contribution to the enhancement of the Josephson current. So we turn to discuss the spatial dependence of the singlet and triplet components on the direction of magnetizations. As shown in Fig.~\ref{Fig.5}, we plot the corresponding singlet component $f_3$ as a function of the coordinate $k_Fy$ for several values of $\theta$. It is found that the amplitudes of $f_3$ appreciably increase with $\theta$ increasing from 0 up to $\pi$. That is because $f_3$ is an even function of $Q$, two singlet components ($f_3^{\rightarrow}$ and $f_3^{\leftarrow}$) originate from left and right superconducting electrodes are nearly symmetrical to each other for different orientations of magnetic moments. From above features we can exclude the contribution of the interference of the singlet component to the long-range proximity effect when the magnetization direction switches from parallel to antiparallel.

  \begin{figure}
  \includegraphics[width=2.9in]{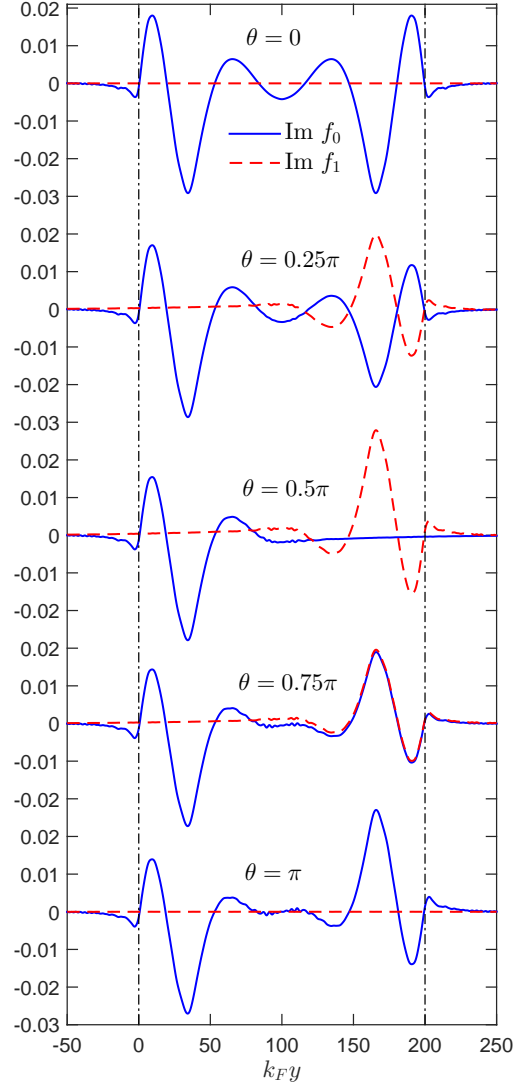} 
  \caption{(Color online) The imaginary parts of triplet components $f_0$ and $f_1$ plotted as a function of the coordinate $k_Fy$ for several values of the misorientation angle $\theta$. Here $k_FL_1$=$k_FL_2$=100, $h_1/E_F$=$h_2/E_F$=0.1, $\omega_Dt$=4, and $\phi$=0. The vertical dash-dotted lines represent the locations of the $S/F_1$ and $F_2/S$ interfaces, respectively.}
  \label{Fig.6}
  \end{figure}

\begin{figure}
  \includegraphics[width=2.9in]{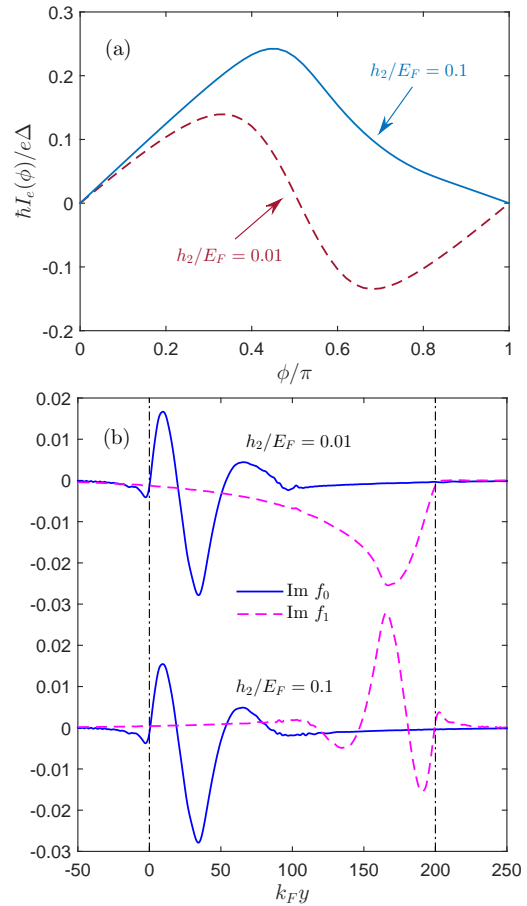} 
  \caption{(Color online) The comparison of asymmetric ($h_1/E_F$=0.1, $h_2/E_F$=0.01) and symmetric ($h_1/E_F$=$h_2/E_F$= 0.1) configurations for the current-phase relation $I_e(\phi)$ (a), and the imaginary part of triplet pair functions $f_0$ and $f_1$ (b) at misorientation angle $\theta$=0.5$\pi$. All results are at fixed values $k_FL_1$=$k_FL_2$=100, $\omega_Dt$=4, and $\phi$=0.}
  \label{Fig.7}
  \end{figure}

  Now let us analyze the dependence of the triplet components on the misorientation angle $\theta$. As illustrated in Fig.~\ref{Fig.6}, for parallel orientation ($\theta=0$) $f_0$ is symmetrical about the center of the $F$ layer. At this time, the equal-spin triplet component $f_1$ does not exist in the entire ferromagnetic region because of the homogeneous magnetization. When the magnetization direction of the $F_2$ layer rotates from the $z$-axis to $x$-axis, the right part of $f_0$ gradually decreases, but $f_1$ correspondingly increases in this region and reaches maximum at $\theta=0.5\pi$. Under this situation, the $F_2$ layer magnetized in the $x$-direction generates the opposite-spin triplet component with respect to the $x$-axis $(\mid\uparrow\downarrow\rangle-$$\mid\downarrow\uparrow\rangle)_{x}$. If one views with respect to the $z$-axis, such state is equivalent to the equal-spin triplet component $-(\mid\uparrow\uparrow\rangle-$$\mid\downarrow\downarrow\rangle)_{z}$~\cite{Esc,MatthiasEschrig}. It is interesting to note that for this perpendicular case the spatial oscillations of $f_0$ in the $F_1$ region will instead exhibit a monotonic spatial variation with jumping into the $F_2$ region. Meanwhile, $f_1$ has the same characteristics as it passes from the $F_2$ layer into the $F_1$ layer. It should be noted that there are two important effects to enhance the supercurrent: (i) the emergence of long-range $f_1$, and (ii) the interference of the $f_0$ and $f_1$. It is well known that $f_1$ could induced in a long range supercurrent. However, if two $F$ layers are highly asymmetric, $f_1$ becomes much larger than $f_0$, then the interference between of them will be reduced accordingly. In this case, the long-range proximity effect manifests itself as a large second harmonic ($I_2\gg{I_1}$) in the spectral decomposition of the Josephson current-phase relation $I(\phi)=I_1\sin(\phi)+I_2\sin(2\phi)+\cdots$. This phenomenon has been proposed in Ref.~\cite{LukaTrifunovic,MengWu}. In contrast, the first harmonic could prevail as the interference of $f_0$ and $f_1$ was restored again in symmetric junction with equal ferromagnetic layers. The comparison of these two cases is shown in Fig.~\ref{Fig.7}. On the other hand, since $\theta$ turns from $0.5\pi$ to $\pi$, $f_1$ gradually decreases but $f_0$ in the $F_2$ region will increase instead, which leads to the enhancement of the interference effect. In the antiparallel configuration $f_1$ completely vanishes, but the interference effect becomes most apparent, which is displayed by the cancellation of $f_0$ in the middle region of the $F$ layer. As a result, in the above process the critical current will continue to increase and reach maximum in antiparallel situation. It is emphasized that the Josephson current in the antiparallel configuration is obviously smaller than that in $SNS$ junctions for the same length between two superconducting electrodes, which has been described in the introduction. That is because the interference effect does not make triplet component $f_0$ cancel out completely in the entire $F$ region, and also does not let the singlet component $f_3$ grow big enough.

\begin{figure}
   \includegraphics[width=3.3in]{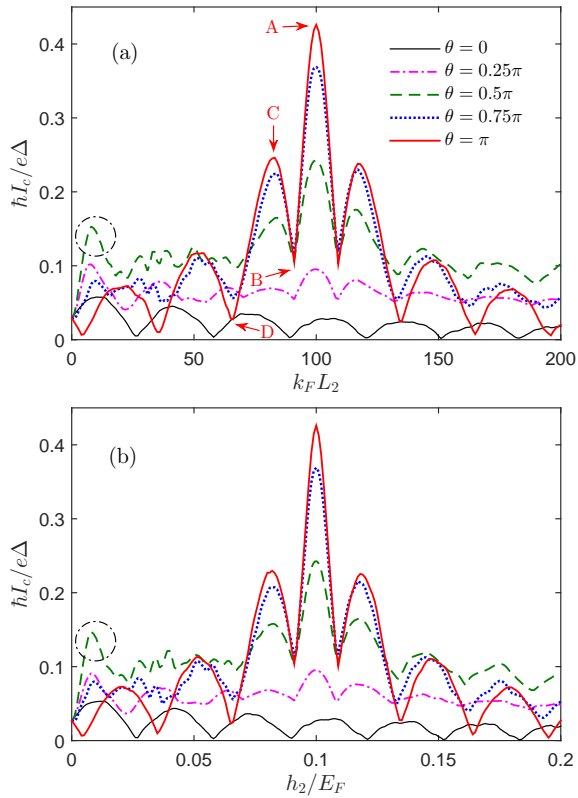} 
   \caption{(Color online) (a) Critical current as a function of thickness $k_FL_2$ for exchange field $h_2/E_F$=0.1. The legend labels the misorientation angles $\theta$. (b) Critical current as a function of $h_2/E_F$ for $k_FL_2$=100. All results are at fixed values $k_FL_1$=100 and $h_1/E_F$=0.1.}
   \label{Fig.8}
   \end{figure}

   To understand further the interference effect of the opposite-spin triplet state, we investigate the intriguing influence of the length and exchange field on the Josephson current when both ferromagnetic layers have different physical features, which is illustrated in Figs.~\ref{Fig.8}(a) and \ref{Fig.8}(b), respectively. Take the first one as an example, the variation of $I_c$ with the thickness $k_FL_2$ looks like a ¡°Fraunhofer pattern¡±. This phenomenon appears more and more obvious as the misorientation angle $\theta$ increases from 0 to $\pi$. In parallel orientation ($\theta=0$), the critical current shows the $0$-$\pi$ conversion on the condition of the nonexistence of interference effect, in which case the amplitude of critical current is weak enough. It is important to note that for perpendicular orientation ($\theta=0.5\pi$) the long range second harmonic current will be induced in highly asymmetric junction, which corresponds to the circular regions denoted in Fig.~\ref{Fig.8}, then the interference effect can be almost negligible. By contrast, $I_c$ dependence exhibits a remarkable oscillating behavior in the thickness range $70<k_FL_2<130$, which marks the enhancement of the interference effect. Moreover, $I_c$ reaches its maximum value for $k_FL_2=100$ and above or below this thickness its amplitude will decrease.

   \begin{figure}
   \includegraphics[width=3.1in]{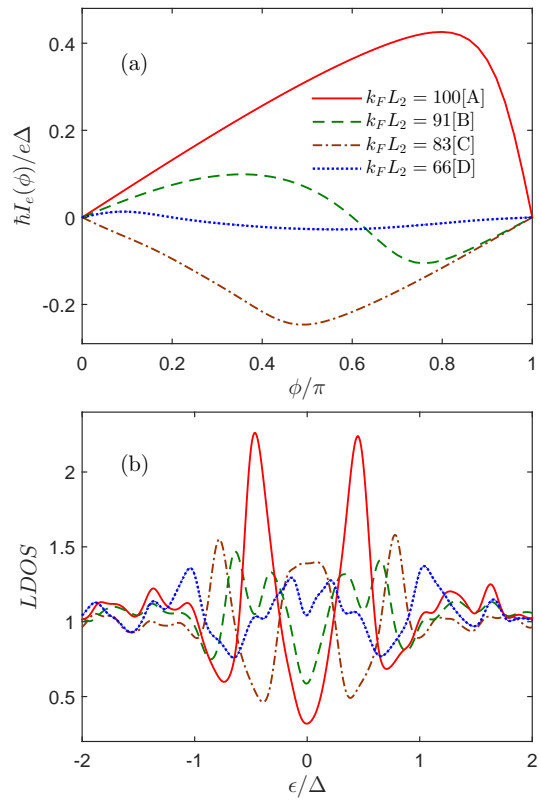} 
   \caption{(Color online) (a) The current-phase relation $I_e(\phi)$ and (b) the LDOS in the center of $F$ layer ($k_Fy$=100) for several thicknesses $k_FL_2$ at misorientation angle $\theta$=$\pi$, which correspond to the point A, B, C and D in Fig~\ref{Fig.8}(a). The LDOS is calculated at $k_BT$=0.0008.}
   \label{Fig.9}
   \end{figure}

   If the two $F$ layers are arranged antiparallel to each other ($\theta=\pi$), the interference effect would appear most likely to occur, meanwhile, their contribution to the Josephson current reaches maximum. In this configuration, we consider in Fig.~\ref{Fig.9} the current-phase relations $I_e(\phi)$ and the corresponding LDOSs in particular points A, B, C and D in Fig.~\ref{Fig.8}(a). If the $F_1$ and $F_2$ layers have identical thickness, as shown in point A, the Josephson current is positive and the LDOS displays a valley at $\epsilon=0$ and two distinguishable peaks at $\epsilon=\pm0.5\Delta$. Besides, when thickness $k_FL_2$ decrease to $91$, corresponding to point B, the Josephson junction is located at the 0-$\pi$ transition point. The first harmonic current vanishes, and the second harmonic will be fully revealed. Subsequently, the sign of $I_c$ turns to negative at $k_FL_2=83$ (point C), and the LDOS at $\epsilon=0$ will be converted from valley to peak. This indicates that the ground state of junction converts into $\pi$ state. At last, the junction will return to critical point of $0$-$\pi$ transition at $k_FL_2=66$ (point D). From Fig~\ref{Fig.8}(a), we can clearly see that the critical current oscillating with $k_FL_2$ displays an unequal period. The detailed explanation will be described in the following paragraph. In addition, if the both $F$ layers have the same lengths but different exchange fields, the critical current $I_c$ shows the similar characteristics (see Fig.~\ref{Fig.8}(b)). This feature illustrates that the interference effect is simultaneously related to the difference of the center-of-mass momenta which are acquired by the spin-opposite triplet pair from the $F_1$ and $F_2$ layers.

\begin{figure}
   \includegraphics[width=2.9in]{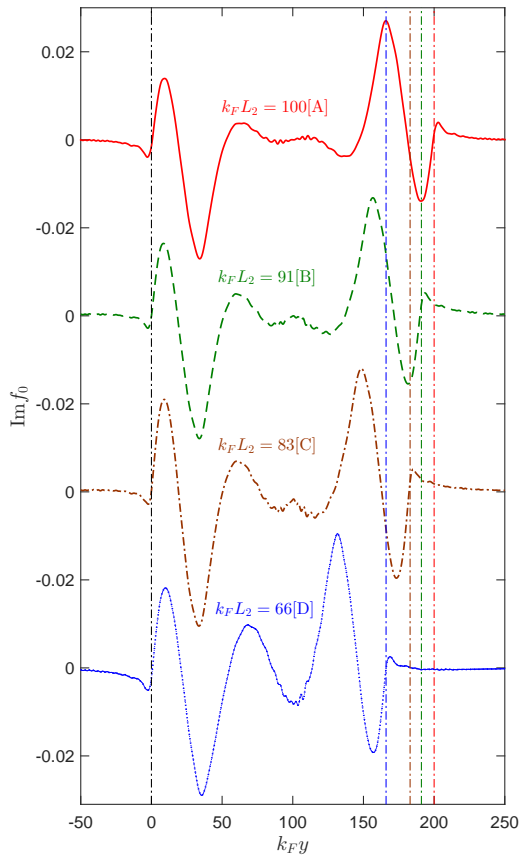} 
   \caption{(Color online) The imaginary parts of triplet pair amplitudes $f_0$ plotted as a function of position $k_Fy$ for several thicknesses $k_FL_2$ at misorientation angle $\theta$=$\pi$, which correspond to the point A, B, C and D in Fig~\ref{Fig.8}(a). Here $k_FL_1$=100, $\omega_Dt$=4, and $\phi$=0. The left vertical dash-dotted line represents the location of the $S/F_1$ interface, and the right ones denote $F_2/S$ interface for different thicknesses $k_FL_2$.}
   \label{Fig.10}
   \end{figure}

   As mentioned before, the interference of $f_0^{\rightarrow}$ and $f_0^{\leftarrow}$ provides the main contribution to the Josephson current. To gain further insight into the interference effect in the asymmetric junctions, we take antiparallel configurations as an example for discussion. In Fig.~\ref{Fig.10}, we present results for the dependence of the triplet components $f_0$ on $k_FL_2$ when the thickness of $F_1$ layer has a fixed value $k_FL_1=100$. It is known that the strength of interference effect is relate to the phase difference and amplitude of two wave functions derived from the opposite direction. We first talk about the contributions of phase difference between $f_0^{\rightarrow}$ and $f_0^{\leftarrow}$ to the oscillation of the critical current. Here we fix the thickness of $F_1$ layer and shorten that of $F_2$ layer, which is similar to set constant $f_0^{\rightarrow}$ and shift $f_0^{\leftarrow}$ from left to right. When both $F$ layers have the same length, the phase difference of two triplet components $f_0^{\rightarrow}$ and $f_0^{\leftarrow}$ is $\pi$ at every position of the $F$ region. In this condition, the interference effect manifests obviously and could induce an enhancement of the Josephson current. As the thickness $k_FL_2$ is reduced to 91, the $f_0^{\leftarrow}$ moves 1/4 period, then the $F_2/S$ interface shift from the red vertical dash-dotted line to the green one. Correspondingly, the junction is situated at the critical point of 0-$\pi$ phase transition. For $k_FL_2=83$, the $f_0^{\leftarrow}$ moves 1/2 period to the right, accordingly, the junction converts to $\pi$ state. Decreasing the $F_2$ layer thickness down to $k_FL_2=66$ means that $f_0^{\leftarrow}$ shifts 3/4 period, the junction return to the critical point of phase transition. It is worth to note that the critical current has unequal oscillation period with varying $k_FL_2$, which is determined by the inhomogeneous spatial oscillation of $f_0^{\leftarrow}$. On other hand, as the length $k_FL_2$ turns from 100 to 0, the mutual cancellation between $f_0^{\rightarrow}$ and $f_0^{\leftarrow}$ will decrease, then the magnitude of $f_0$ will be enhanced by the superposition of above both triplet components. This phenomenon indicates a weakening of the interference effect that can make the Josephson current diminish.

   \section{Conclusion}

    In this paper, we have investigated the relationship between the long-range Josephson current and the pairing correlations in clean $SF_1F_2S$ junctions with the misorientation magnatizations through solving the BdG equations. The interference effect of the opposite-spin triplet component was pointed out as a source of this current. The main reason is because the Josephson critical current will enhance when the magnetizations rotate from the parallel to the antiparallel orientation. In this process, the singlet component changes slightly but the interference effect of the triplet components $f_0^{\rightarrow}$ and $f_0^{\leftarrow}$ will increase correspondingly, and in the antiparallel configuration the interference of both components cloud nearly cancel each other in central ferromagnetic region. This behavior can be attribute to two facts: (i) the triplet components $f_0^{\rightarrow}$ and $f_0^{\leftarrow}$ are derived from the $S/F_1$ and $F_2/S$ interfaces and transmit to opposite directions. They experience a scattering occurred at the $F_1/F_2$ interface and take this interface as an emission source to continually spread into another ferromagnetic layer. (ii) The antiparallel magnetizations will provide opposite center-of-mass momentum to the Cooper pair, then two singlet components $f_3^{\rightarrow}$ and $f_3^{\leftarrow}$ almost maintains invariant, but the triplet components $f_0^{\rightarrow}$ and $f_0^{\leftarrow}$ have opposite sign and could interference cancellation in the $F$ region. In addition, if the feature of the $F_1$ layer remain unchanged, the interference effect will make the critical current oscillate with the length and exchange field of the $F_2$ layer. Therefore this finding provides new insight into the physical mechanism to the long-range proximity effect in the Josephson junctions with non-parallel magnetizations and can be important for the implementation of interference effect in superconducting spin electronic devices.

   \section*{Acknowledgments}

   This work was supported by the National Natural Science Foundation of China (Grants No.11447112, No.51106093, No.11547161, No.11547039 and No.11504222), the Scientific Research Program Funded by Shaanxi Provincial Education Department (Grants No.15JK1132, No.12JK0972, No.15JK1150 and No.15JK1111), the Opening Project of Shanghai Key Laboratory of High Temperature Superconductors (Grant No.14DZ2260700), and the Scientific Research Foundation of Shaanxi University of Technology (Grants No.SLG-KYQD2-01, No.SLGKYQD2-02 and No.SLGKYQD2-03). J. Wu would like to thank the Shenzhen Peacock Plan and Shenzhen Foundamental Research Foundation (Grant No.JCYJ20150630145302225).

    \end{document}